\documentclass[pdflatex,sn-mathphys]{sn-jnl}
\jyear{2025}
\usepackage{caption}
\usepackage{multibib}
\newcites{supp}{Methods References}
\usepackage{afterpage}
\usepackage{rotating}
\usepackage{amsmath}

\usepackage{lineno}
\modulolinenumbers[1]
\begin{document}

\title{Star formation in the circumgalactic high-velocity cloud Complex H}

\author[1]{\fnm{Zhihong} \sur{He}}

\author[1]{\fnm{Wenkang} \sur{Pang}}

\author[1]{\fnm{Kun} \sur{Wang}}

\author[1]{\fnm{Yangping} \sur{Luo}}

\author[1]{\fnm{Qian} \sur{Cui}}

\affil[1]{\orgdiv{School of Physics and Astronomy}, \orgname{China West Normal University}, \orgaddress{\city{Nanchong}, \country{China}}. Corresponding author: Z.He \url{hezh@mail.ustc.edu.cn}}

\maketitle
\vspace{-1cm}


\section*{Abstract}\label{sec:Abstract}
\textbf{The accretion of metal-poor gas sustains galactic star formation. In the Milky Way, this process is fueled by high-velocity clouds (HVCs), yet their fundamental properties have remained elusive in the absence of stellar tracers. Here we report a binary open cluster within HVC Complex~H. With an age of 11.2 $\pm$ 0.6\,Myr and a subsolar metallicity of 0.05$^{+0.05}_{-0.02}$\,Z$_\odot$, the clusters provide a direct stellar distance anchor to the cloud at 13.8 $\pm$ 0.6\,kpc. Their proper motions indicate Complex~H is on a prograde, south-to-north orbit through the outer Galactic disk. The resulting interaction produces a ``slow-fast-slow'' velocity gradient, with the cloud's outer layers decelerating as they merge into the disk. Orbit integration suggests the clusters formed from an internal cloud-cloud collision. This triggering mechanism implies other HVCs could similarly produce high-velocity stars. The scarcity of previous stellar detections in HVCs is explained by the rapid escape of young stars ($<$ 20\,Myr), while CO non-detections may stem from weak emission due to low metallicity and gas dispersal. This work reveals that the circumgalactic medium can sustain star formation, offering a tangible laboratory to probe the physical conditions of accreting gas before it merges with the Galactic disk.}


HVCs are neutral hydrogen (HI) structures first identified in the 1960s through 21 cm emission \citep{Muller1963}. They are frequently detected in Galactic HI surveys \citep{Stark1992,Barnes2001,Kalberla2005,Winkel2016} and are commonly defined as gas with $\mid v_{\mathrm{lsr}}\mid > 90$ km s$^{-1}$. Those exceeding $\mid v_{\mathrm{lsr}}\mid > 170$ km s$^{-1}$ are classified as very high-velocity clouds (VHVCs). Such extreme velocities cannot be explained by Galactic rotation models. Consequently kinematic distance estimation methods become unreliable for them. The lack of any detected stellar counterparts or star formation signatures in HVCs has prevented direct measurement of their distances, metallicities, and dynamics \citep{Siegel2005,Simon2006,Stark2015}. Absorption-line spectroscopy toward background stars can offer distance or metallicity constraints in some sightlines \citep[e.g.,][]{vanWoerden1999, Wakker2001}, but many regions do not contain suitable background sources, especially for VHVCs \citep{Lehner2022}. These persistent limitations have led to continued debate about the origins and evolutions of HVCs \citep[e.g.][]{Oort1970,Bregman1980,Lu1998,Fox2023,Lucchini2024}. 

The HVC known as Complex H exemplifies these challenges. Located in the second Galactic quadrant, it exhibits multi-phase neutral gas \citep{Lockman2003,Kalberla2006} with line-of-sight velocities ranging from -220 to -120 km s$^{-1}$. Its distance remains poorly constrained: absorption-line spectra of foreground stars indicate a lower limit beyond 5 kpc \citep{Wakker1998}, while kinematic models suggest it may lie as far as approximately 27 kpc, potentially on a retrograde orbit \citep{Lockman2003}. Given its low Galactic latitude and complex velocity structure, Complex H is hypothesized to be interacting with the outer Galactic disk \citep{Morras1998}. Some studies further support this scenario by proposing that a collision with outer Galactic disk gas at a distance of $\sim$16 kpc may have triggered star formation within the disk clouds \citep{Izumi2014, Izumi2024}. Yet, as with other HVCs, the absence of stellar populations within Complex H itself \citep{Simon2006} has precluded definitive confirmation of its physical connection to the Milky Way.
﻿
\section*{Results}\label{sec:Results}

\begin{figure}
    \centering
	\includegraphics[width=1.\columnwidth]{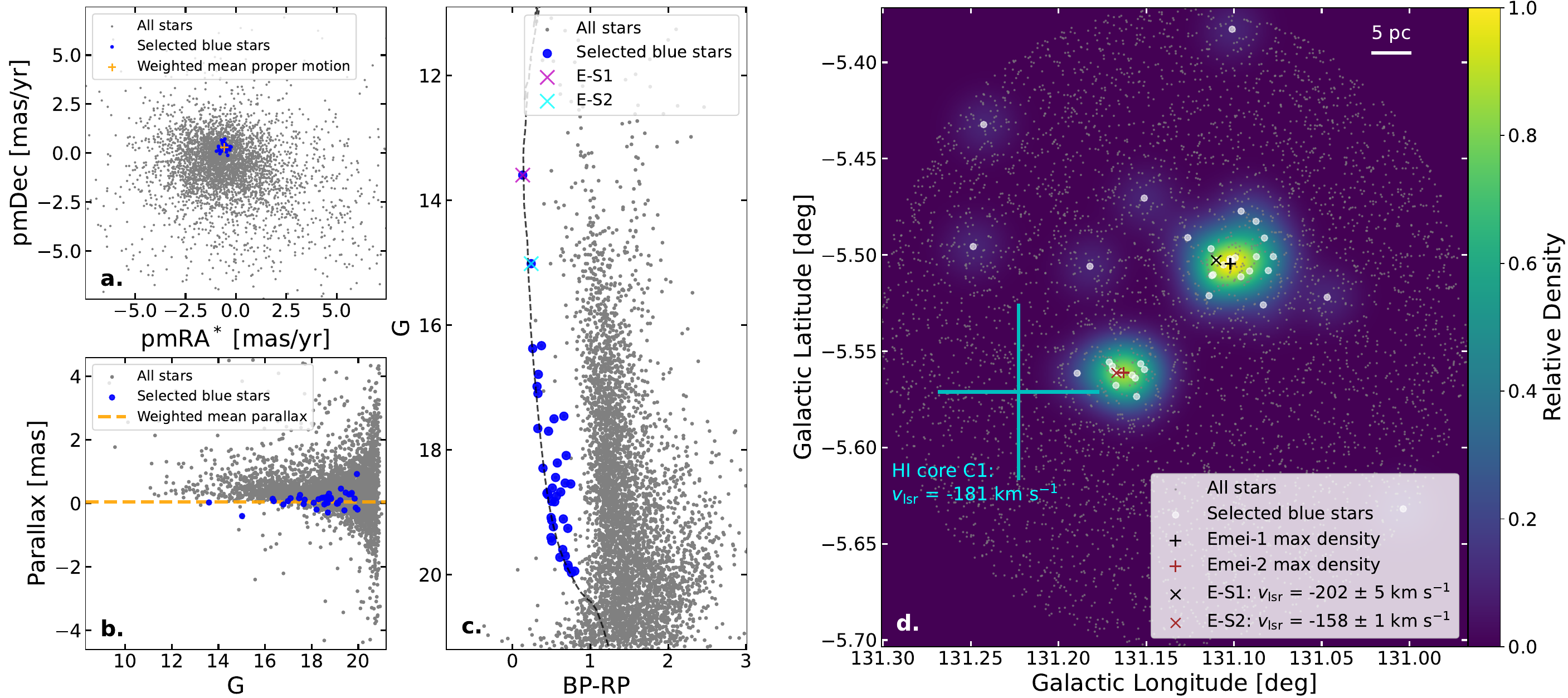}
\caption{\textbf{Gaia detection of young binary cluster in cold cloud C1.} The figure presents astrometric and photometric properties of the Emei star clusters and their massive members (E-S1 and E-S2). (a) Proper motion distribution of candidate member stars. The cluster exhibits tightly grouped kinematics with a mean proper motion ($\bar{\mu}_{\mathrm{\alpha^*}}$, $\bar{\mu}_\mathrm{\delta}$) = (-0.58 $\pm$ 0.01, 0.30 $\pm$ 0.02) mas yr$^{-1}$. (b) Magnitude versus parallax for the same stars, confirming a common distance with a mean parallax $\bar{\varpi}$ = 0.05 $\pm$ 0.02 mas. (c) Colour–magnitude diagram showing a clear blue main sequence. The distribution matches an isochrone-derived age of 11.2 Myr at a distance of 13.8 kpc (black dashed line), which corresponds to a Galactocentric radius of 20.0 kpc. Cluster members are provided in Extended Data Fig.~2. (d) Spatial density distribution of those blue stars, revealing two compact and distinct clusters Emei-1 and Emei-2. The scale bar corresponds to 5 pc; and the cyan cross marks the pixel position of the peak HI emission in C1, with the arm lengths representing the maximum sampling interval of the EBHIS survey.
}
    \label{fig:Figure_1}
\end{figure}

Fortunately, this long-standing issue can now be addressed. Using astrometry from Gaia \citep{GaiaCollaboration2021}, spectroscopy from Large Sky Area Multi-Object Fiber Spectroscopic Telescope (LAMOST) \citep{Deng2012,Zhao2012}, and HI cloud data from the Effelsberg-Bonn HI Survey (EBHIS) \citep{Winkel2016}, we have discovered a pair of young open clusters, designated Emei-1 and Emei-2, spatially and kinematically coincident with Southern Complex H (SCH, see Methods and Extended Data Fig. 1). As shown in Fig.~\ref{fig:Figure_1}, the cluster members share consistent proper motions and parallaxes. Their color-magnitude diagrams reveal prominent blue sequences that are clearly distinct from the field population. The blue stars form two spatially compact groups separated by approximately 0.1$^\circ$ (Extended Data Fig. 2), indicating a binary open cluster structure. The clusters lie very close to the peak emission of cold core C1 ($v_\mathrm{lsr}$ = -181 km s$^{-1}$) in SCH, with angular separations of only 0.05$^\circ$ and 0.15$^\circ$, respectively. These offsets are comparable to the EBHIS maximum sampling interval ($0.09^{\circ}$), corresponding to a projected physical distance of less than $\sim 20$\,pc.

LAMOST spectroscopic observations are available for the brightest star in each cluster (E-S1 in Emei-1 and E-S2 in Emei-2; see Methods and Extended Data Fig.~3). Both are classified as early B-type stars. The spectra yield $v_\mathrm{lsr}$ values of -202 $\pm$ 5 km s$^{-1}$ for E-S1 and -158 $\pm$ 1 km s$^{-1}$ for E-S2, giving a mean systemic $v_\mathrm{lsr}$ of -180 km s$^{-1}$ \citep{Xiang2022}. The derived metallicities are similarly subsolar, with $Z_\mathrm{E-S1}$ = 0.04$^{+0.06}_{-0.02}$ $Z_\odot$ and  $Z_\mathrm{E-S2}$ = 0.06$^{+0.07}_{-0.03}$ $Z_\odot$ \citep{Xiang2022}, consistent with the low-metallicity regime observed in other HVCs \citep[e.g.,][]{vanWoerden1999, Wakker2001}. These velocities greatly exceed those of the farthest known disk gas ($v_\mathrm{lsr}$ $\sim$ -100 km s$^{-1}$ in HI and CO emission \citep{Izumi2014,Digel1994}) and match the VHVC component of SCH. As shown in Fig.~\ref{fig:Figure_1} and Extended Data Fig.~2, the two massive stars lie near the center of each cluster. E-S1, in the richer cluster Emei-1, is more luminous than E-S2 in Emei-2, consistent with the higher number of member stars in Emei-1.

Isochrone fitting (Methods) yields a young age of 11.2 $\pm$ 0.6 Myr for both clusters, consistent with the spectral classification of E-S1 and E-S2 as B-type main-sequence stars. The derived distance moduli indicate a common distance of 13.8 $\pm$ 0.6 kpc. Although the individual stellar velocities differ by about 20 km s$^{-1}$ from the peak HI velocity in C1, their mean value agrees well with the gaseous component. This spread in stellar velocities may originate from binary orbital motion and/or measurement uncertainties. 
Collectively, the high radial velocities and strong spatial association firmly link the clusters to the VHVC.

\begin{figure}
    \centering
	\includegraphics[width=.95\columnwidth]{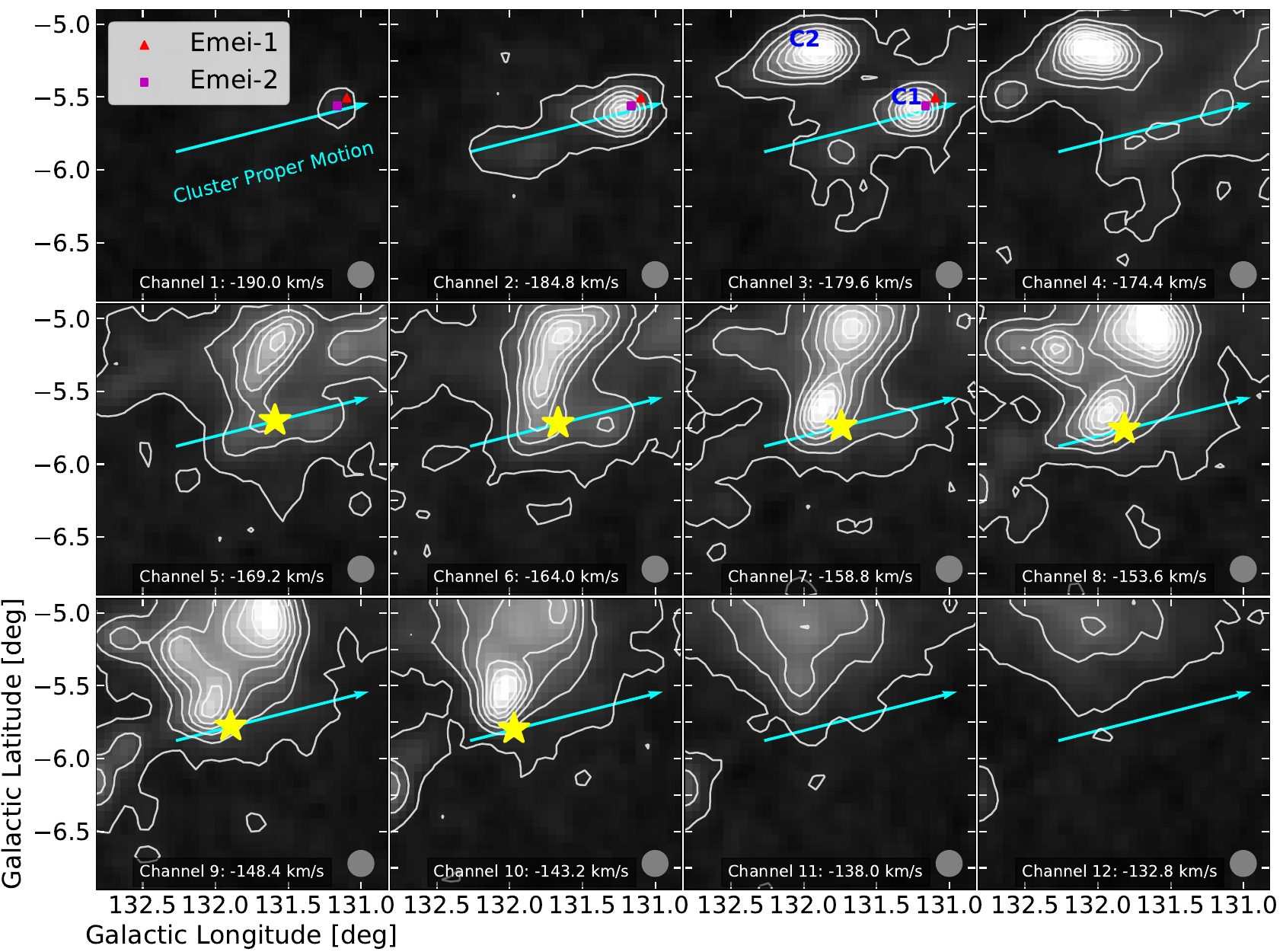}
  \caption{\textbf{HI channel maps of the HVC complex surrounding the Emei clusters.} Each panel shows the neutral hydrogen distribution within a different velocity channel, with the central $v_\mathrm{lsr}$ indicated at the bottom. The grey circles show the angular resolution of the EBHIS survey. Cluster positions are marked by triangle (Emei-1) and square (Emei-2). Contours start at 1 K km s$^{-1}$ and increase in steps of 2 K km s$^{-1}$. The contours in Channel 2 exhibit a clear cometary morphology. An arrow indicates the clusters' proper motion direction, which passes through the center of C1. The proper motion is closely aligned with the elongation of the head-tail structure. This distinctive geometry is also visible for core C2 and across Channels 2 to 4. Star symbols mark the present-day locations of potential progenitor cloud candidates at different $v_{\mathrm{lsr}}$ ranges that may have collided with C1 11.2 Myr ago. They are closely associated with a dense clump in SCH at ($l,b$ $\sim$ 132.0$^{\circ}$, -5.6$^{\circ}$) within SCH.}
    \label{fig:Figure_2}
\end{figure}

To examine the local environment of the young binary clusters in greater detail, we analyzed HI channel maps covering their vicinity. As shown in Fig.~\ref{fig:Figure_2}, the cold gas component, characterized by narrow velocity dispersions ($\sigma_v$ $\sim$ 2 km s$^{-1}$), is concentrated in two compact structures: C1 and C2, both detected in the highest-velocity channels (Channels 2 to 4; $v_\mathrm{lsr}$ between -190 and -175 km s$^{-1}$). In contrast, the warm gas, with broader velocity dispersions ($\sigma_v$ $\sim$ 15 km s$^{-1}$), is distributed over a wider velocity range ($v_\mathrm{lsr}$ $\lesssim$ -180 km s$^{-1}$).

Both C1 and C2 exhibit cometary morphologies with head-tail structures that share a consistent orientation (Channels 2 to 4). C1 situated at higher radial velocities and farther from the diffuse warm emission, displays a clearer and more extended structure with minimal blending. In Channel 2, at a contour level of 1 K km s$^{-1}$, the feature extends $1.33^{\circ} \times 0.45^{\circ}$, which corresponds to an intrinsic physical size of $\sim 320 \times 99$\,pc after beam deconvolution. The Emei clusters are located precisely at the head of this structure. Importantly, the direction of the tail extension aligns closely with the proper motion vectors of the clusters. This geometric and kinematic coherence indicates that the cloud core likely experienced compression along the direction of motion \citep{Bruns2000,Putman2011}.
Orbit analysis provides further evidence that C1 collided with a proximate, dense gas clump within SCH (Methods and Fig. \ref{fig:Figure_2}). Such an internal cloud-cloud collision may have triggered the formation of the young clusters in this HVC.

\begin{figure}
    \centering
	\includegraphics[width=1.\columnwidth]{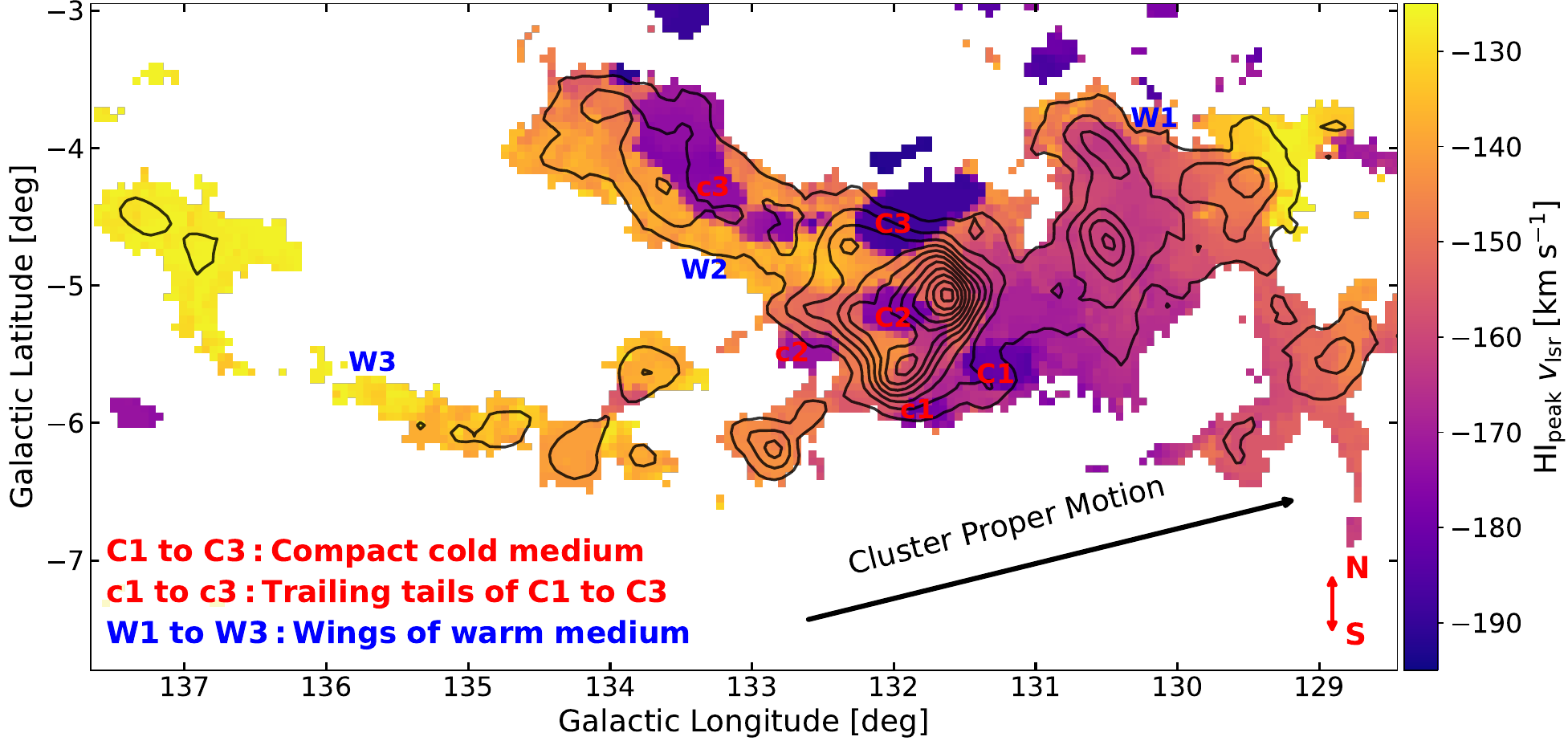}
\caption{\textbf{Kinematic structure of SCH.} Color map indicates the velocity at the peak intensity in the high-velocity component, including only pixels with peak brightness temperature above 5$\sigma_\textrm{rms}$ (Methods). Black contours represent the integrated intensity of the warm gas ($v_\mathrm{lsr}$ = -170 to -120 km s$^{-1}$), starting from 15 K km s$^{-1}$ in steps of 10 K km s$^{-1}$. The upward/downward arrow indicates the direction toward the north/south Galactic pole. Prominent wings are visible: W1-W3 originate in the warm gas and exhibit velocities lower than the main body, while the northern cold component C3 also shows an elongated, lagging tail oriented similarly to the wings. In contrast, cold cores C1 and C2 display narrow tails aligned with the direction of motion, suggesting a different formation mechanism.}
    \label{fig:Figure_3}
\end{figure}

On a larger scale, the velocity structure of SCH (Fig. \ref{fig:Figure_3}) reveals a compact and fragmented configuration, with its major axis closely aligned with the proper motion of the young clusters, suggesting a common dynamical origin. Additionally, SCH exhibits northward-extending wings that show significant velocity deceleration ($\Delta v_\mathrm{lsr}$ $>$ 30 km s$^{-1}$), which may be attributed to gas stripping by Kelvin-Helmholtz instabilities \citep{Murray1993,Barger2020,Sander2021}, with potential contributions from other mechanisms \citep[e.g.][]{Armillotta2017,Shelton2022}. 
More importantly, the velocity field reveals a systematic gradient along the direction of motion. The central regions of SCH show the most negative velocities around $-175$~km~s$^{-1}$, while both the leading and trailing edges exhibit slower motions at approximately $-130$~km~s$^{-1}$. 
This combined geometric and kinematic pattern aligns with gas cloud interactions \citep[e.g.][]{Benjamin1997,Marchal2021}, confirming that SCH is being shaped by its environment.

\section*{Discussion}\label{sec:Discussion}

We then modeled the trajectory of cloud core C1 while accounting for drag forces from the ambient medium (Methods and Extended Data Fig.~4). Our analysis indicates that an upper limit on the ambient disk gas density is required to explain why C1 remains spatially associated with the young clusters despite the ongoing interaction. This implied density is on the order of $10^{-24}\,\text{g cm}^{-3}$ (Extended Data Fig.~5), which corresponds to the lower end of the range expected for the diffuse warm neutral medium \citep{Marchal2021,Audit2005}. Such conditions support the ``peloton effect'' scenario \citep{Forbes2019,Heitsch2022}, where trailing gas becomes shielded from ram pressure and may eventually overtake the HVC's leading portions.

The entire Complex H shows a similar ``slow-fast-slow'' (SFS) pattern in latitude-velocity space. This structure extends approximately 100 km s$^{-1}$ in velocity and spans $15^{\circ}$ in latitude (Methods and Extended Data Fig.~6). Previous studies have interpreted this SFS feature as evidence of retrograde infall from north to south at a distance of 27 kpc \citep{Lockman2003}. In that model, the velocity gradient across the main body of Complex H ($-1^{\circ} < b < 5^{\circ}$) is attributed to geometrical projection effects. In contrast, our analysis supports a prograde interaction scenario in which Complex H moves from south to north into the Galactic disk. This interpretation is consistent with the systematic velocity pattern observed in SCH.

Meanwhile, the south-to-north infall scenario better matches the observed distance of the young clusters and aligns with the warped structure of the outer Galactic disk, which appears elevated in the northern region at $l \approx 130^{\circ}$ where Complex H is located \citep[e.g.][]{Levine2006,Skowron2019,Chen2019,He2023}. We find that Complex H's velocity structure smoothly merges with the disk gas at both northern and southern extremities (Methods), indicating the cloud's outer layers are blending into the disk component.


Orbital integration (Methods and Extended Data Fig.~7) further indicates that the Emei clusters will enter high-inclination orbits and cross the disk in the next $\sim$30 Myr. Within this period, both E-S1 and E-S2 are expected to undergo supernova explosions due to their massive B-type nature. These explosions are likely to reshape the morphology of the outer disk gas and potentially trigger new star formation in the Outer Scutum-Centaurus arm. Separately, the impact of Complex H appears to have already induced star formation in the disk \citep{Izumi2014, Izumi2024}. 
Over time, tidal forces from the Galactic potential are expected to dissipate the Emei clusters, possibly leading to a fate similar to that of known loose stellar associations in the halo. Price-Whelan 1 represent one such example \citep{PriceWhelan2019,Nidever2019}. In contrast to the Emei clusters, it is older ($\sim$100~Myr), has already crossed the disk toward higher latitudes, and has separated from its proposed gaseous progenitor. However, it shares key characteristics with the Emei system---notably subsolar metallicity and high $v_{\mathrm{lsr}}$ implying that this system may also have formed via an HVC (likely the Magellanic Leading Arm \citep{Nidever2019}).

The discovery of the Emei clusters within SCH provides an important opportunity to study star formation in near-pristine, low-metallicity environments and suggests that some hypervelocity stars and streams may originate in HVCs. Confirming such origins remains difficult, however, as stars born in HVCs quickly ($< 20$ Myr, see Methods) become kinematically detached from their parent gas through the complex dynamics \citep{Marchal2021,Armillotta2017,Shelton2022,Barger2020} of the ambient medium. This rapid decoupling also accounts for the absence of older stars and molecular line emission in previous surveys of Complex H \citep{Simon2006}. When gas-rich systems such as dwarf galaxies are accreted by the Milky Way, stars and giant molecular structures experience less ram pressure deceleration than the diffuse gas, leading them to separate over time from the more extended low-density components.

Only very young stellar groups retain unambiguous kinematic and spatial association with their birth clouds. In addition, the low metallicity of HVCs makes traditional molecular tracers like CO inherently faint \citep{Bell2006,DessaugesZavadsky2007}. Observationally, detecting diffuse CO emission is further complicated when off positions contain line signals. Deep, targeted observations of dense, cold HI cores may therefore offer a more effective strategy for identifying high-velocity molecular gas (Methods). The current non-detections of CO in HVCs likely arise from these combined observational limitations. Nevertheless, the formation of the Emei clusters strongly suggests that dense molecular gas could indeed be present within shielded regions of Complex H.


\section*{Methods}\label{sec:method}

\subsection*{HI survey data and reduction}
The core region of Complex H was first identified in the 1970s \citep{Hulsbosch1971,Dieter1971} and later confirmed through HI surveys and high-resolution observations to extend over 15 degrees \citep{Wakker1991,Buenrostro2008}. We analyzed the 21 cm neutral hydrogen emission using data from the EBHIS \citep{Winkel2016}, which is well-suited for this study due to its full northern sky coverage and favorable angular resolution $\theta_{\mathrm{beam}}$ = 10.8 arcmin. The EBHIS data have a spectral resolution of 1.49 km s$^{-1}$ and a typical rms noise level of $\sigma_\mathrm{rms}$ = 90 mK, covering a velocity range of -600 to 600 km s$^{-1}$ in the local standard of rest frame.

We produced an integrated intensity map (Extended Data Fig.~1) by summing the brightness temperature over the velocity interval from -220 to -170 km s$^{-1}$, which encompasses the very high-velocity cloud component of Complex H. For comparison, we also generated a map over the lower-velocity range from -170 to -120 km s$^{-1}$ and overlaid it as contours. Previous studies have mostly concentrated on the northern part of the complex (HVC131+1-200), while its southern extension has received little attention. Extended Data Figure~1 shows that the VHVC emission ($v_\mathrm{lsr}$ = -200 to -170 km s$^{-1}$) in Complex H is predominantly located at Galactic latitudes above $b = -3^\circ$, with only a few compact cores observed at lower latitudes. 

Within Complex H, the lower-velocity HVC emission ($v_\mathrm{lsr}$ = -170 to -120 km s$^{-1}$) exhibits a more diffuse distribution and displays a prominent structure near $(l, b)$ = (131.5$^\circ$, -5$^\circ$), extending approximately 2 degrees. This structure has a large velocity dispersion of about 15 km s$^{-1}$, suggesting it belongs to the warm neutral medium. In contrast, the adjacent VHVC cores show low velocity dispersion of approximately 2 km s$^{-1}$, characteristic of the cold neutral medium (see the sample spectrum in Extended Data Fig.~1 and the channel map in Fig.~2). This two-phase nature is also observed in the main body of Complex H (at $b > -3^\circ$), as previously reported \citep{Lockman2003,Kalberla2006}. The spatial proximity, kinematic coherence, and shared multi-phase structure indicate that the VHVC cores and the lower-velocity HVC emission form a single complex. Furthermore, the young binary clusters Emei-1 and Emei-2 are located precisely at the position of the compact VHVC core HVC131.2-5.6-181. No other HI emission features are detected in coincidence with the clusters.

To investigate the spatial and kinematic relationship between the stellar clusters and their associated gas, we analyzed the velocity structure using spectral channel maps. The channel maps in Fig.~2 were constructed with a velocity resolution of 4$\delta v$, where $\delta v$ = 1.29 km s$^{-1}$ corresponds to the original channel width of the EBHIS data (not the spectral resolution). We also generated a peak velocity map (Fig.~3) that displays for each pixel the $v_\mathrm{lsr}$ value corresponding to the maximum brightness temperature in the high-velocity component of SCH. To reduce contamination from noise, pixels with peak intensities below 0.5 K (5$\sigma_\textrm{rms}$) were masked.

To examine the broader spatial scale, Extended Data Figure~6 presents six position-velocity diagrams constructed from EBHIS data across Galactic longitudes 134$^{\circ}$ to 128$^{\circ}$, with each panel displaying the emission structure in latitude-velocity space for a specific 1$^{\circ}$ longitude interval. The diagrams consistently reveal a characteristic SFS pattern throughout Complex H. While ref. \cite{Lockman2003} previously identified this kinematic structure near $l \sim 130.5^{\circ}$, our extended coverage shows that at both latitudinal extremities of the SCH gas appears progressively closer in velocity to the disk gas. This velocity convergence suggests that the head and tail of Complex H may be actively merging into the Galactic disk.

For clarity, we refer to HVC131.2-5.6-181 as C1, where the nomenclature HVC+$l$+$b$+$v_\mathrm{lsr}$ conventionally denotes the Galactic longitude, latitude, and central $v_\mathrm{lsr}$ of the HVC. Similarly, we designate HVC131.9-5.2-176 and HVC131.9-4.5-187 as C2 and C3, respectively. C1 to C3 represent the only three compact cold components clearly distinguishable from the warmer gas in velocity space within SCH. Their fainter, trailing cold gas structures are labeled c1 to c3 (Fig.~3).


\subsection*{Identification of young clusters in HVC}
We developed a method to identify stellar counterparts of HVCs based on the idea that if star formation occurs in compact HVCs, young stars are expected to remain near their birth sites. Young stellar populations may still be spatially and kinematically associated with the gas, as they have not had sufficient time to disperse fully. This expectation is supported by observations of Galactic spiral arms, where clusters younger than 20 Myr are often found near their birthplaces \citep[e.g.,][]{Dias2005}.

Confirming whether individual stars originated within HVCs is observationally challenging, as firmly establishing their ages and natal clouds is often difficult. OB stars are exceptions due to their youth and clear spectral features. We therefore focused on identifying groups of young stars such as OB associations or young open clusters, which are more likely to exhibit measurable astrometric and photometric coherence. Since OB stars typically form in clustered environments \citep{McKee2003,Zinnecker2007}, their presence serves as a reliable tracer of recent vigorous star formation. The detection of a stellar aggregate containing OB stars near a compact HVC thus offers stronger evidence for in situ star formation than single stars. Our search specifically targeted such systems in the vicinity of HVCs.

To identify such stars, we used the hot star catalog from the LAMOST survey \citep{Xiang2022}. This catalog is based on spectra obtained with the 4-meter LAMOST telescope \citep{Zhao2012,Deng2012}, which has a low-resolution (R $\sim$ 1800, $\lambda$ $\sim$ 3800-9000 $\textrm{\AA}$) spectra, corresponding to a velocity resolution of about 70 km s$^{-1}$. We selected main-sequence stars with high radial velocities ($\mid \textrm{RV}\mid$ $>$ 150 km s$^{-1}$ ) and high effective temperatures ($T_\textrm{eff}$ $>$ 20 000 K). Additional constraints included a signal-to-noise ratio $>$ 50 and surface gravity in the range 4 $<$ log g $<$ 5. These criteria produced only 23 candidates across the entire sky, highlighting the rarity of young ($<$ 30 Myr), very high-velocity massive OB stars. 

Among the selected candidates, two sources stand out due to their exceptional properties. Designated E-S1 and E-S2, these objects correspond to LAMOST spectral IDs 20151101-HD020325N544136V01-16-225 and 20151101-HD020325N544136B01-16-219, respectively. They are both the hottest sources in the sample and form the only binary pair, with an angular separation of merely 0.1$^\circ$. All other candidates are isolated. Furthermore, both stars exhibit spatial and kinematic coincidence with the cloud core C1 in SCH. Spectral analysis confirms their classification as early B-type stars, displaying strong hydrogen and helium absorption lines (Extended Data Fig.~3). E-S1 and E-S2 emerged as the most promising candidates based on their high temperatures, binary nature, and positional alignment with SCH. To assess physical association, we further examined their kinematic properties using Gaia DR3 astrometry \citep{GaiaCollaboration2021}.

The Gaia space observatory provides high-precision astrometry at the ten-microarcsecond level, enabling accurate parallax and proper motion measurements for over a billion stars in the Milky Way \citep{GaiaCollaboration2016}. In the region surrounding the high-velocity stars E-S1 and E-S2, we identified a distinct blue sequence in the color-magnitude diagram (Fig.~1), indicative of a young stellar population. These blue stars exhibit clustering in both astrometric parameters and spatial distribution. We therefore applied a rigorous member selection process using 3 sigma clipping in proper motion and parallax, along with a color selection of BP-RP $<$ 0.8 to isolate the bluest stars.

Although these selection criteria do not completely eliminate field star contamination, the filtered sample shows a well-defined sequence in both the color-magnitude diagram and the vector point diagram (Fig.~2). The observed dispersions in proper motion and parallax also match those of known young open clusters in the Galactic disk \citep[e.g.,][]{CantatGaudin2020,He2023b}. This astrometric coherence, together with the photometric properties, strongly suggests that the blue stars form a genuine stellar cluster system in an early evolutionary phase.

To assess the statistical significance of the clustering, we performed a surface density analysis using a Gaussian kernel estimation method \citep{Virtanen2020}. The resulting density map (Fig.~1) shows two clear peaks near the positions of E-S1 and E-S2. Each peak corresponds to a cluster core with a central density that exceeds the density of the field background by more than a factor of 10. The two clusters are clearly separated. The spatial distribution of each cluster is highly compact. Most member stars lie within a radius of 0.01$^\circ$ to 0.02$^\circ$ , equivalent to 2.5 to 5 pc at the estimated distance of 13.8 kpc. This tight spatial structure together with the kinematic coherence provides strong evidence that both are gravitationally bound systems. 
A cross-match with current open cluster catalogs \citep{He2023b,HR2023} shows that these systems are not included in existing catalogs.


\subsection*{Photometric data reduction}
We analyzed the color–magnitude diagrams of both clusters using PARSEC isochrones \citep{Bressan2012} (Extended Data Fig.~2). Potential systematics in Gaia BP-band photometry were considered, which may overestimate fluxes for faint red stars but have negligible effect on blue stars \citep{Riello2021}. Consistency checks using other optical bands (e.g., Pan-STARRS \citep{Chambers2016}) confirmed that this effect does not impact the member stars in this study. We determined the value and uncertainties in age and extinction by visually inspecting the fit of the isochrones to the blue edge of the member star distribution.

Fits with solar metallicity models did not match the observed sequences. This discrepancy arises because metallicity significantly affects the location of the main-sequence turn-on in young clusters; therefore, satisfactory agreement was only achieved using subsolar metallicities with [M/H] $<$ -0.6. The derived distance modulus is highly sensitive to the assumed metallicity. For example, a metallicity of [M/H] = -1 corresponds to a distance modulus of 15.8 (corresponding to a distance of 14.5 kpc), while [M/H] = -2 gives 15.5 (12.6 kpc). Notably, this metallicity-dependent distance variation does not affect the main conclusions of this study. In contrast, the estimated age and extinction remain robust across this metallicity range. The best-fitting isochrones give log(age/yr) = 7.05 for both clusters, with an uncertainty of 0.025 dex.

The extinction values show a small but significant difference between the two clusters. From our isochrone fitting, Emei-1 exhibits $A_\textrm{V}$ = 1.25 $\pm$ 0.05 mag, while Emei-2 shows slightly higher extinction of $A_\textrm{V}$ = 1.35 $\pm$ 0.05 mag. These values, derived from the blue edge of the main sequence in our fitting, are slightly larger than foreground extinction estimates of 1.02 $\pm$ 0.06 mag obtained through multi-wavelength photometry from ref.\cite{Schlafly2011,Green2019}. This slight excess suggests additional extinction from dust local to their birth environment. Additionally, although most stars align well with the isochrone, some deviate from the binary sequence, suggesting the presence of differential extinction within the region. For this well-defined population, we estimate a total cluster mass of $2.1 \pm 0.6 \times 10^3~M_\odot$ by comparing the observed luminosity function to theoretical isochrones, following the method of ref.\cite{Jadhav2021,Cui2025}. This total comprises $1.2 \pm 0.5 \times 10^3~M_\odot$ for Emei-1 and $0.9 \pm 0.2 \times 10^3~M_\odot$ for Emei-2. We note that this is a preliminary estimate that does not yet account for the effects of unresolved binaries or differential extinction.

Based on the mean metallicity of [M/H] = -1.3 $\pm$ 0.3 from LAMOST spectroscopic measurements of E-S1 and E-S2 \citep{Xiang2022}, we derived a distance modulus of 15.7 $\pm$ 0.1, corresponding to a heliocentric distance of 13.8 $\pm$ 0.6 kpc. This photometric distance is slightly smaller than that derived from the weighted mean Gaia DR3 parallax of 53 $\pm$ 18 $\mu$as ($18.9^{+9.7}_{-4.9}$ \,kpc). However, distance measurements from parallax at this distance range contain significant uncertainties. These include a zero-point error of approximately -20 $\mu$as \citep{Lindegren2021} that currently limits the accuracy of purely astrometric distances. After accounting for this zero-point offset, the distance derived from parallax agrees with the photometric distance from the isochrone-based modulus.

The youth of these clusters indicates that they likely remain near their birth site within the HI cloud core, which would have been decelerated by ram pressure from the surrounding gaseous medium. Their clear detection by Gaia and LAMOST suggests that stellar feedback has already dispersed much of the original dust, while the young age of the system implies that no supernova explosion has yet disrupted the cloud or triggered significant cluster dispersal.

\subsection*{Mass Estimation of C1}
The neutral hydrogen mass of C1 was derived from the EBHIS spectral lines. As shown in Fig.~\ref{fig:Figure_2}, the head-tail structure that contains C1 has an observed width of $0.45^{\circ}$ (2.5~$\theta_{\mathrm{beam}}$). Accounting for beam convolution, the intrinsic angular width is $\theta_{\mathrm{real}} = \sqrt{\theta_{\mathrm{obs}}^2 - \theta_{\mathrm{beam}}^2} \approx 0.41^{\circ}$. At a distance of 13.8\,kpc, this corresponds to a physical width of $\sim 99$\,pc. We integrated the HI emission of C1 over the velocity range of $[-185.0, -176.5]$\,km\,s$^{-1}$. At an integrated intensity contour of 5\,K\,km\,s$^{-1}$, the core exhibits a circular morphology with a physical diameter of approximately 80\,pc ($\theta_{\mathrm{cloud}} \approx 20.4'$). It is worth noting that the physical size of the EBHIS beam corresponds to $\sim 43$\,pc at this distance. Consequently, compact cores smaller than the sampling limit, if present, would likely be heavily diluted. Due to the lack of higher-resolution HI data or complementary molecular observations (e.g., CO) to trace dense gas, our current analysis considers only the diffuse component resolved by EBHIS, without accounting for potential unresolved dense clumps.

The HI mass was calculated using the standard formulation for optically thin gas:

\begin{equation}
M_{\mathrm{HI}} = 0.046 \times \langle W_{\mathrm{HI}} \rangle \times r^2 \times f_{\mathrm{beam}}
\end{equation}
where $\langle W_{\mathrm{HI}} \rangle$ is the average integrated intensity in K km s$^{-1}$, $r$ is the cloud radius in parsecs, $f_{\mathrm{beam}}$ is the beam dilution correction factor. The coefficient 0.046 incorporates fundamental constants and unit conversions, derived from:
\begin{equation}
0.046 = \frac{m_{\mathrm{H}}}{M_{\odot}} \times (1.823 \times 10^{18}) \times \pi \times (3.086 \times 10^{18})^2
\end{equation}
where $m_{\mathrm{H}} = 1.674 \times 10^{-24}$ g is the hydrogen atom mass, $M_{\odot} = 1.989 \times 10^{33}$ g is the solar mass, and the numerical factors account for unit conversions between cm$^2$ and pc$^2$.

A beam dilution correction factor of $f_{\mathrm{beam}} = 1 + (\theta_{\mathrm{beam}} / \theta_{\mathrm{cloud}})^2 \approx 1.28$ was applied.
With an average integrated intensity of $\langle W_{\mathrm{HI}} \rangle = 21.9$ K km s$^{-1}$ measured across 31 pixels within the circular region, the HI mass was determined to be $2.2 \times 10^3~M_\odot$. This calculation yields specifically the neutral hydrogen mass. To estimate the total gas mass including helium, we applied a standard correction factor of 1.4, giving a total gas mass of approximately $3.1 \times 10^3~M_\odot$.

Notably, the presence of young clusters implies that molecular hydrogen must have been present, as star formation typically occurs in molecular clouds. This suggests that mass of core C1 was dominated by H$_2$. 
To provide a rough estimate of the gas reservoir, we adopt a characteristic star formation efficiency (SFE) of 10$\%$. We emphasize that this value is illustrative, serving only to facilitate a qualitative assessment of the progenitor cloud rather than a precise mass determination.
Based on this assumption and our estimated cluster mass, we infer a total progenitor cloud mass of $\sim 2 \times 10^4~M\odot$. 
While SFE values can vary, 10$\%$ is often cited as typical for star-forming cores in cloud-cloud collision scenarios \citep[e.g.][]{Higuchi2010,Chen2024}. Assuming no significant mass loss, this would imply a mass for C1 of approximately $1.8 \times 10^4~M_\odot$. 
We caution that the subsequent dynamical analysis and molecular gas estimates for C1 depend critically on this assumed SFE and should be interpreted as tentative constraints.

\subsection*{Dynamics of cloud core}

We analyze the dynamical evolution of the cloud core C1 over the past 11.2 Myr, corresponding to the current age of the stellar clusters. Setting $t_0 = 0$ as the formation time of the clusters, the cloud core had an initial radius $r_0$ and was moving with velocity $v_0$ relative to the ambient interstellar medium (ISM).

Consider the cloud radius increases linearly with time at a rate $k_0$, such that $r_t = r_0 + k_0 t$. Based on the current radius of $\sim 40$ pc and an assumed initial radius of $r_0 = 5$ pc, we derive $k_0 = 3.1$ km s$^{-1}$. For a uniformly expanding sphere, this expansion velocity corresponds to a line-of-sight velocity dispersion of $\sigma_v \approx k_0/\sqrt{3} = 1.8$ km s$^{-1}$, slightly smaller than the observed velocity dispersion of 2.1 km s$^{-1}$ in the core center, suggesting predominantly coherent expansion with mild internal turbulence.

Assuming constant mass $m$ for the cloud core, it experiences a drag force while moving through the ISM of density $\rho$. The drag force follows the expression $F = -\frac{1}{2} C \rho A v^2$ \citep{Tan2023}, where $C$ = 0.5 is the drag coefficient for sphere and $A = \pi r_t^2$ is the cross-sectional area, respectively. This leads to a drag acceleration $a_t = -k_1 r_t^2 v_t^2$, with $k_1 = \frac{1}{2} C \rho \pi / m$. The effects of Galactic gravity are neglected in this analysis, as they act equally on both the cloud core and the ambient ISM and do not significantly affect their relative motion over this timescale.

At $t_0 = 0$, we assume the cloud core formed the star clusters with zero initial separation between them. The clusters continue to move with the constant initial velocity $v_0$, while the cloud core experiences deceleration due to drag. To determine the cloud core velocity $v_t$ after time $t$, we solve the differential equation derived from the drag acceleration:

\begin{equation}
\frac{dv}{dt} = -k_1 (r_0 + k_0 t)^2 v^2
\end{equation}

Assuming the initial velocity \(v_0\) at $t_0$ and solving the present velocity \(v_t\):

\begin{equation}
v_t = \frac{1}{\frac{1}{v_0} + \frac{k_1}{3k_0} \left[ (r_0 + k_0 t)^3 - r_0^3 \right]}
\end{equation}

The separation $s_t$ between the cloud core and star cluster after time $t$ is calculated as the difference between the distance traveled by the cluster ($v_0 t$) and the distance traveled by the cloud core:

\begin{equation}
s_t = v_0 t - \int_0^t v(t) dt
\end{equation}

We now proceed to numerical simulations based on this framework. Given that the velocity difference between the cloud core and the ambient medium is primarily in the vertical direction ($36$ km s$^{-1}$) and along the line of sight ($75$ km s$^{-1}$), with minimal proper motion difference in the tangential direction (0.02 mas yr$^{-1}$, corresponding to $1.3$ km s$^{-1}$), we applied the model to fit both the velocity and spatial offsets in $v_z$ and $v_{\text{lsr}}$ (Extended Data Fig.4).

The simulation adopts a cloud core mass of $1.8 \times 10^4\ M_\odot$, derived assuming a SFE of 10$\%$ for C1. Observationally, the SFE in star-forming molecular cores typically ranges from about 3$\%$ to 30$\%$ \citep[][]{Higuchi2010}. This implies that the estimated mass of C1 could vary by up to a factor of three. Such quantitative uncertainty does not alter the qualitative conclusions of our subsequent analysis.

As shown in Extended Data Fig.~4, The velocity reduction $\Delta v$ as a function of ISM density after 11.2 Myr was evaluated for initial velocities of 40 km s$^{-1}$ and 95 km s$^{-1}$, respectively. The results suggest that the $v_{\mathrm{lsr}}$ of the clusters may have been approximately -200 km s$^{-1}$. Furthermore, the density-dependence of the vertical velocity ($v_z$) constrains the upper limit of the ambient medium density to 0.5 to $1 \times 10^{-24}$ g cm$^{-3}$, such that the resulting vertical displacement remains less than about 15 pc (Fig.~1). 
This value is already below the EBHIS sampling scale ($\sim$22 pc at the distance of 13.8 kpc) and current observations cannot resolve structures smaller than the beam size, the true physical separation may be significantly smaller. Therefore, our derived displacement serves strictly as a conservative upper limit.
On the other hand, if the actual mass of C1 were higher than current estimates (i.e., with SFE $< 10\%$), the upper density limit would shift toward the higher end of this range. This density is characteristic of, though slightly lower than, typical diffuse warm HI gas \citep[e.g., $\sim 1 \times 10^{-24}$ g cm$^{-3}$ in ref.][]{Marchal2021,Audit2005}, indicating that C1 may still be shielded by the main body of SCH and thus encountering a relatively low-density environment.

Over the next 5 to 10 Myr, the vertical separation between C1 and the clusters is expected to accelerate, leading to their disengagement from the cloud core in the vertical direction. In contrast, along the line of sight, the clusters likely became detached from C1, which could explain the lack of infrared \citep{Wright2010} or radio continuum \citep{Stein2021} emission detected from this cloud core; however, due to their similar proper motions, this has not resulted in a significant tangential separation.

\subsection*{Cluster orbital investigation}
To trace the dynamical trajectory of the young star clusters, we simulated their orbital evolution using a Python package $Galpy$ \citep{Bovy2015} and the Galactic gravitational potential model MWPotential2014 \citep{Bovy2013}, which is designed for Galactic dynamics studies. We adopted the following solar parameters from ref.\cite{Reid2019}: a Galactocentric distance $R_0$ = 8.15 kpc, a circular velocity $\Theta_0$ = 236.7 km s$^{-1}$, and solar peculiar motion relative to the local standard of rest ($U_\odot$, $V_\odot$, $W_\odot$) = (10.6, 10.7, 7.6) km s$^{-1}$. The resulting orbital trajectory is presented in Extended Data Fig.~7.

We converted $v_\mathrm{lsr}$ = -181 km s$^{-1}$ of C1 to heliocentric radial velocity, yielding RV = -185 km s$^{-1}$. To express proper motions in Galactic coordinates, we transformed the mean proper motion of the Emei clusters ($\mu_{\mathrm{\alpha^*}}$, $\mu_\mathrm{\delta}$) = (-0.58, 0.30) mas yr$^{-1}$ to ($\mu_{\mathrm{l^*}}$, $\mu_\mathrm{b}$) = (-0.63, 0.16) mas yr$^{-1}$. We adopted a heliocentric distance of 13.8 kpc from isochrone fitting. These values correspond to a current position of $R$ = 20.0 kpc from the Galactic center and a velocity of ($v_\mathrm{r}$, $v_\mathrm{\phi}$, $v_\mathrm{z}$) = (-69, -201, 36) km s$^{-1}$ for the core C1. The rotation curve show a $v_{\mathrm{lsr}} = -105$ km s$^{-1}$ for the objects at this radius ($v_{\mathrm{\phi}} = -224$ km s$^{-1}$ in ref.\cite{Reid2019}).

\subsection*{Present-day position of collision companion cloud}

The young clusters likely formed following C1's collision with a potential dense clump along its trajectory. To locate the present-day position of this collision site, we investigate the spatial separation between C1 and clump candidates at different $v_{\mathrm{lsr}}$.

We first exclude disk gas as the colliding material because its proper motion in Galactic longitude ($\mathrm{pml^*} = -0.65$ mas yr$^{-1}$) closely matches that of C1 ($\mathrm{pml^*} = -0.63$ mas yr$^{-1}$), and we find no dense cloud cores near $v_{\mathrm{lsr}} = -105$ km s$^{-1}$ in the region around C1. Instead, the collision likely involved dense gas within SCH itself, with $v_{\mathrm{lsr}}$ between $-170$ and $-140$ km s$^{-1}$ (Fig.~2).

The velocity difference between these dense clumps and the disk gas is smaller---only one to two-thirds of C1's initial velocity difference. Since the drag force scales with the square of velocity difference, these slower clumps experience deceleration reduced by half to an order of magnitude compared to C1. We therefore treat the dense clump as moving at constant velocity relative to C1.

We then calculate the positional differences in Galactic longitude ($dl$) and latitude ($db$) over 11.2 Myr. Our analysis focuses on drag-induced changes in C1's $v_{\mathrm{lsr}}$ and proper motion. In Galactocentric coordinates, disk gas has rotational velocity $v_\phi = -224$ km s$^{-1}$ \citep{Reid2019}, close to C1's $v_\phi = -201$ km s$^{-1}$. The resulting drag produces negligible change in $v_\phi$ (only 1-2 km s$^{-1}$), allowing us to treat it as constant. As $v_z$ changes do not affect $\mathrm{pml^*}$, and assuming the collision occurs along C1's proper motion direction, we set $db = dl \times (\mathrm{pmb}_{\mathrm{C1}}/\mathrm{pml^*}_{\mathrm{C1}})$, where $\mathrm{pmb}_{\mathrm{C1}}$ and $\mathrm{pml^*}_{\mathrm{C1}}$ are C1's proper motion components. Extended Data Figure~5 (a) and (b) show the evolution of C1's $v_{\mathrm{lsr}}$ and corresponding proper motion under a density of $0.5 \times 10^{-24}$ g cm$^{-3}$, respectively.

The angular separation in longitude after 11.2 Myr is:
\begin{equation}
dl = \int_0^{11.2\ \mathrm{Myr}} \mathrm{pml^*}_{\mathrm{C1}}(t)  dt - \mathrm{pml^*}_{\mathrm{Clump}} \times 11.2\ \mathrm{Myr} \quad
\end{equation}
where the first term integrates C1's time-varying proper motion and the second assumes constant proper motion for the dense clump. 
As shown in Extended Data Fig.~5, the simulated positional differences between C1 and clumps with different $v_{\mathrm{lsr}}$ range from 0.4 to 0.8 degrees, with clumps lagging behind C1 as Galactic longitude decreases. These results remain largely unchanged even when increasing the ISM density by a factor of two.


\subsection*{Potential molecular component in C1}

We estimate the current CO emission level in C1 by assuming a SFE of 10$\%$. After subtracting the known atomic hydrogen and helium masses, the remaining molecular gas mass is $M_{\mathrm{H_2}} \approx 1.0 \times 10^4~M_{\odot}$,  with a full width at half maximum (FWHM) line width of $\Delta v \approx 4$~km~s$^{-1}$, comparable to the observed HI line width (or maybe a little smaller). Based on this mass, we calculate both the average integrated intensity and radiation temperature of the CO spectral line using standard methods involving the gas surface density, H$_2$ column density, and the CO-to-H$_2$ conversion factor ($X$).

The gas surface density was determined as:
\begin{equation}
\Sigma_{\mathrm{gas}} = \frac{M}{\pi R^2} = \frac{1.0 \times 10^4 ~\mathrm{M_{\odot}}}{\pi \times (40 ~\mathrm{pc})^2} \approx 4.1 \times 10^{-4}~\mathrm{g}~\mathrm{cm}^{-2}
\end{equation}

The H$_2$ column density was then calculated using:
\begin{equation}
N(\mathrm{H}_2) = \frac{\Sigma_{\mathrm{gas}}}{2~m_{\mathrm{H}}} \approx 1.2 \times 10^{20}~\mathrm{cm}^{-2}
\end{equation}

The $X$-factor was determined to be:
\begin{equation}
X = k_X \times X_{\odot} = 2k_X \times 10^{20}~\mathrm{cm}^{-2}~(\mathrm{K}~\mathrm{km}~\mathrm{s}^{-1})^{-1}
\end{equation}
where $k_X$ represents the ratio of the CO-to-H$_2$ conversion factor relative to the solar neighborhood value $X_{\odot}$, which varies with metallicity and is typically enhanced in low-metallicity environments \citep{Bell2006,Bolatto2013}.

The average integrated intensity of the spectral line was calculated as:
\begin{equation}
W(\mathrm{CO}) = \frac{N(\mathrm{H}_2)}{X} = \frac{0.6}{k_X}~\mathrm{K}~\mathrm{km}~\mathrm{s}^{-1}
\end{equation}

The average radiation intensity (brightness temperature) was derived from the integrated intensity and FWHM:
\begin{equation}
T_{\mathrm{peak}} = \frac{W(\mathrm{CO})}{\Delta v} = \frac{0.15}{k_X}~\mathrm{K}
\end{equation}

Given that the CO-to-H$_2$ ratio is metallicity-dependent and can be substantially elevated in low-metallicity environments (below 0.3 - 0.5 Z$_\odot$ \citep{Bolatto2013}). The expected CO line intensity may be therefore be as low as a few mK or lower. This low predicted intensity helps explain the previous non-detection of CO in Complex H \citep[][]{Simon2006}. Furthermore, if the actual star formation efficiency were higher than the 10$\%$ adopted here, even less molecular gas would remain in C1, making its CO emission correspondingly fainter and more challenging to detect.

However, denser cores that are currently forming or have not yet expanded and dispersed may exist within Complex H. For instance, a compact core of a few parsecs ($\sim 1$ arcmin at 13.8 kpc) containing a molecular mass similar to that of C1 could produce a CO line intensity on the order of hundreds of mK. Emission from such compact regions would likely be far easier to detect than the diffuse emission from C1.

Simultaneously, high-resolution HI observations are also indispensable to resolve the fine-scale neutral gas structure, which is vital for probing the physical link and properties of the gas associated with these stellar clusters. While the EBHIS data utilized in this work provide the highest resolution currently available for Complex H, they remain limited in resolving such small-scale structures as core C1, preventing a more precise quantitative analysis of the star-gas interaction at this stage.



\section*{Data Availability}
All datasets analysed in this study are available from public online archives. The astrometric data were accessed through the Gaia Archive \citep{GaiaCollaboration2016, GaiaCollaboration2021} at \url{https://gea.esac.esa.int/archive/}. The spectra of cluster member stars E-S1 and E-S2 were obtained from the LAMOST \citep{Deng2012,Zhao2012} data release at \url{https://www.lamost.org/lmusers/}. The EBHIS \citep{Winkel2016} HI data are available at \url{http://cdsarc.u-strasbg.fr/ftp/J/A+A/585/A41/fits/}, and PARSEC \citep{Bressan2012} isochrones can be downloaded from \url{https://stev.oapd.inaf.it/cgi-bin/cmd}. The object IDs and processed data products are available from $zenodo$ \citep{ZenodoDataHe2026}.

\section*{Code Availability}

The code for $Astropy$ \citep{Astropy2013,Astropy2022} is publicly available at \url{https://docs.astropy.org/en/stable/index.html}. The $Galpy$ \citep{Bovy2015} code can be accessed at \url{https://www.galpy.org/}. Code generating all figures is available from $zenodo$ \citep{ZenodoDataHe2026}.


\section*{Acknowledgements}
This work was supported by National Natural Science Foundation of China (NSFC) through grants 12573023,12303024, the “Young Data Scientists” project of the National Astronomical Data Center (NADC2023YDS-07), and the Natural Science Foundation of Sichuan Province (2024NSFSC0453); K.W. is supported by NSFC12373035; Y.L. is supported by the NSFC12573035,12173028, and the China Space Station Telescope project: CMS-CSST-2021-A10. The two star clusters presented in this work are identified as Emei-1 and Emei-2. The identifier acknowledges the Emei Mountain, a cultural landmark in Sichuan, China. The Gaia image in Extended Data Fig. 7 is used under the ESA Standard Licence. And this work has made use of data from the European Space Agency mission Gaia, processed by the Gaia Data Processing and Analysis Consortium (DPAC). Funding for the DPAC has been provided by national institutions, in particular the institutions participating in the Gaia Multilateral Agreement.

\section*{Author Contributions}
Z.H. conceived the study, conducted the data analysis, wrote the manuscript, and revised it in response to the reviewers' comments. 
W.P. and Q.C. assisted in the preparation of the Extended Data Fig.~7 and reference formatting.
K.W. and Y.L. discussed the interpretation of the LAMOST spectra. 

\section*{Competing interests}

The authors declare no competing interests.

\clearpage

\clearpage

\clearpage

\newpage
\captionsetup[table]{name=Extended Data Table}
\captionsetup[figure]{name=Extended Data Figure}

\setcounter{figure}{0}


\begin{figure}
\center
	\includegraphics[width=1\columnwidth]{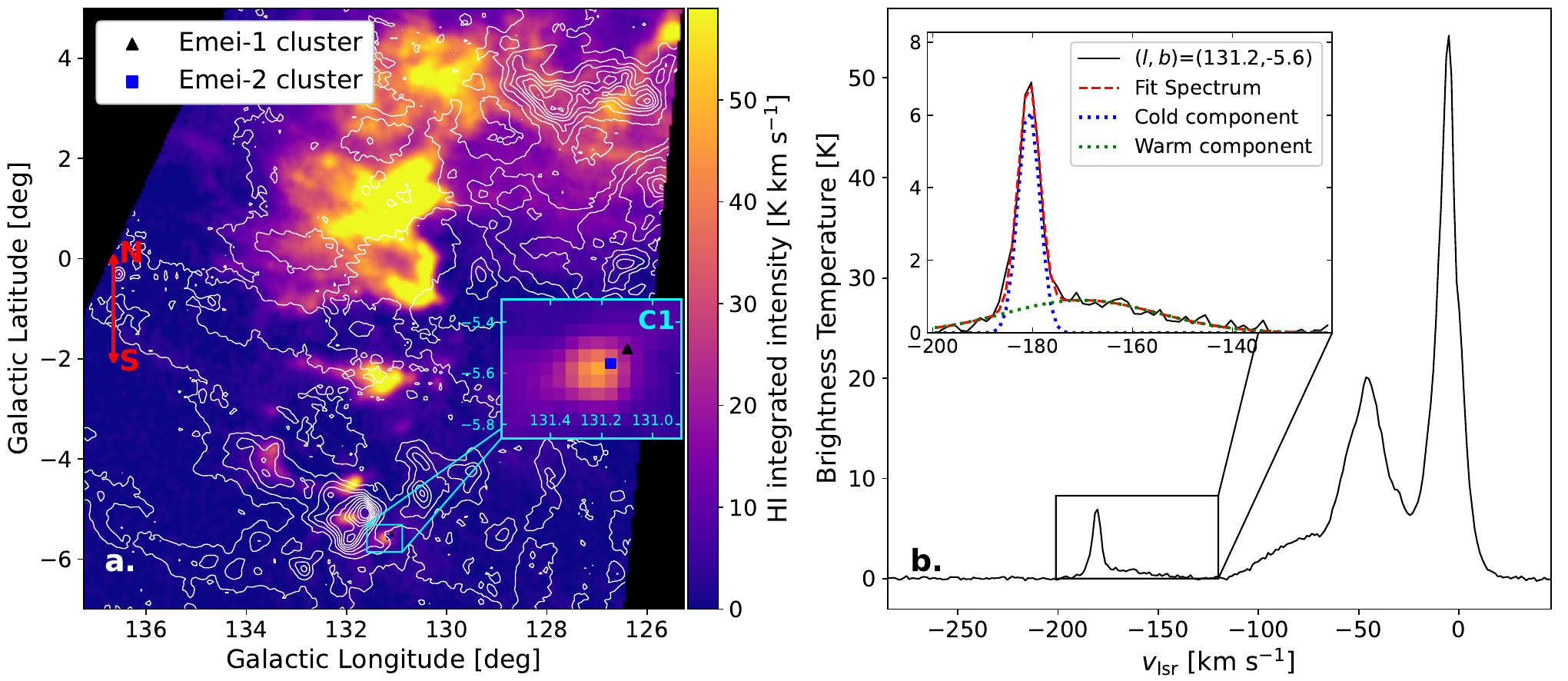}
\caption{\textbf{Neutral hydrogen emission from Complex H in EBHIS.} (a) Integrated intensity map of HI emission from the VHVC component ($v_\mathrm{lsr}$ = -220 to -170 km s$^{-1}$; color scale). Contours show the lower-velocity component ($v_\mathrm{lsr}$ = -170 to -120 km s$^{-1}$), starting at 5 K km s$^{-1}$ with steps of 10 K km s$^{-1}$. The upward/downward arrow indicates the direction toward the north/south Galactic pole. The inset offers a detailed view of the cold compact core C1 in SCH, which is spatially and kinematically linked to the young clusters Emei-1 and Emei-2. (b) HI spectrum at the peak emission pixel of C1. The inset highlights the high-velocity emission ($v_\mathrm{lsr} < -120$  km s$^{-1}$) with a Gaussian fit, revealing both warm (velocity dispersion $\sigma_v$ = 15.2 km s$^{-1}$) and cold ($\sigma_v$ = 2.1 km s$^{-1}$) neutral gas components. These kinematic properties are characteristic of Complex H \citep{Lockman2003,Kalberla2006}.}
    \label{fig:Extended_Data_Figure_1}
\end{figure}

\clearpage

\begin{figure}
\center
	\includegraphics[width=1\columnwidth]{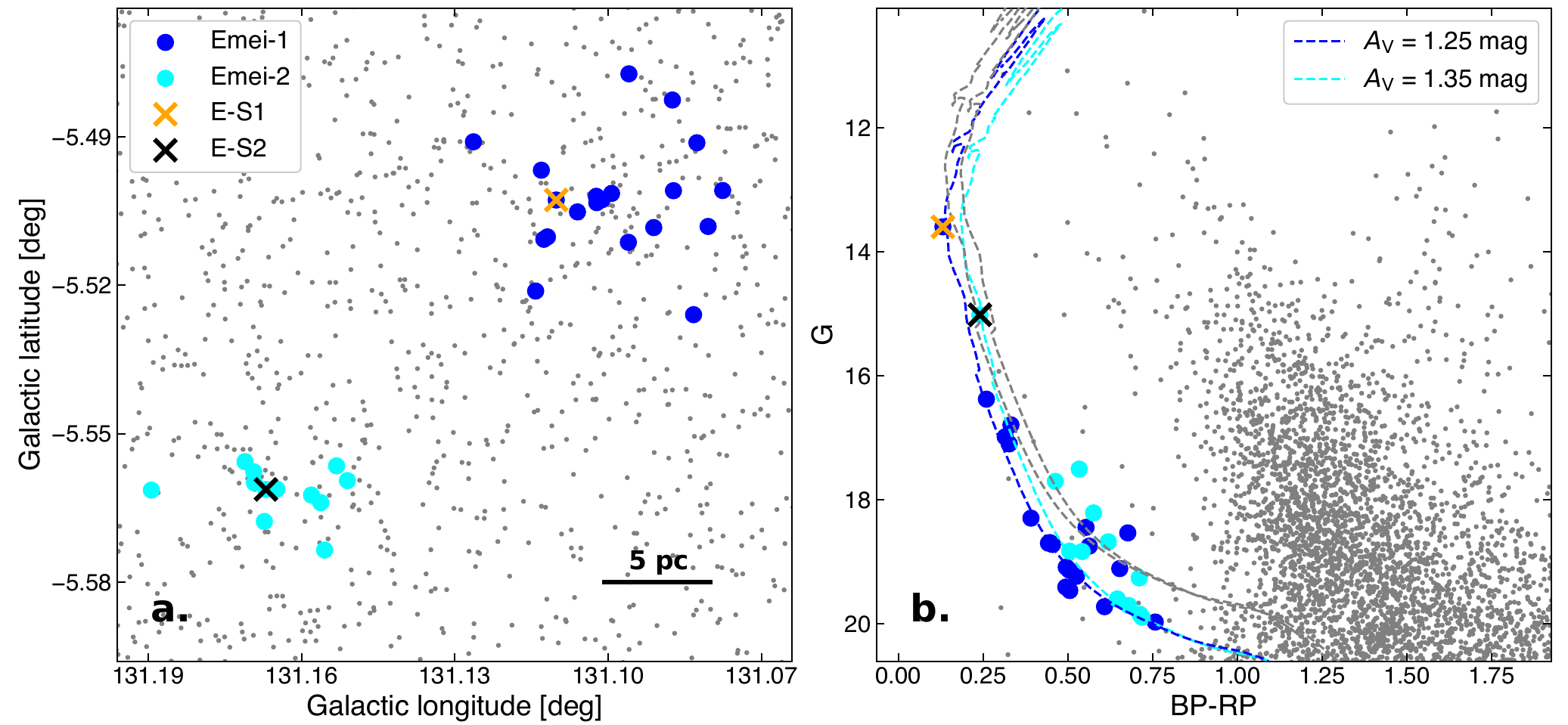}
   \caption{\textbf{Isochrone fitting for each cluster.} (a) Spatial distribution of member stars in Galactic coordinates for Emei-1 (blue) and Emei-2 (cyan), showing compact and concentrated stellar populations. The scale bar corresponds to 5 pc. The two clusters are spatially separated by approximately 20 pc. (b) Color-magnitude diagram displaying the main sequences of both clusters. Isochrone fits (dashed lines) correspond to an age of log(age/yr) = 7.05. Emei-2 exhibits higher extinction than Emei-1, consistent with its closer proximity to the dense cloud core C1. Grey dashed lines indicate possible binary sequences. Some stars deviate from the best-fit isochrones, likely due to significant differential extinction within the cloud core causing variations in local extinction.}
    \label{fig:Extended_Data_Figure_2}
\end{figure}

\clearpage

\begin{figure}
\center
	\includegraphics[width=1\columnwidth]{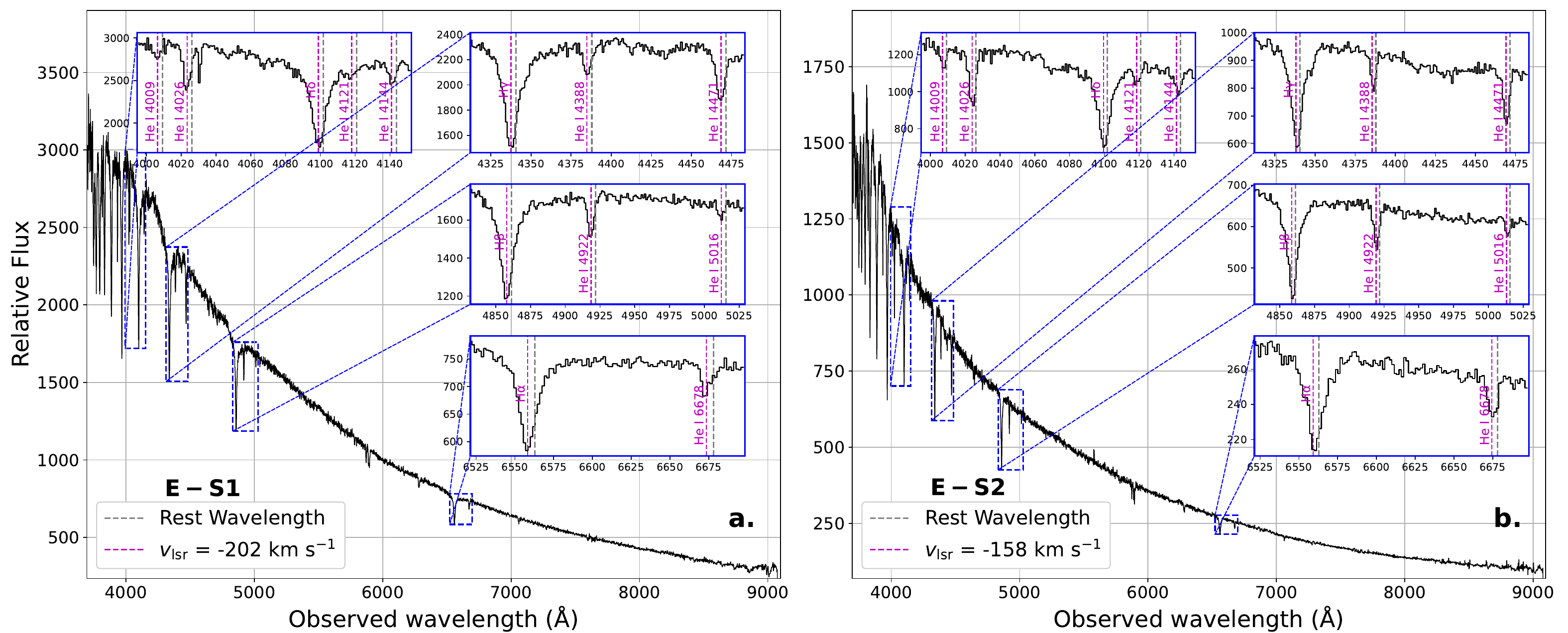}
\caption{\textbf{Spectroscopic characterization of the two massive young stars in SCH.} The observed LAMOST spectra (black step curves) exhibit characteristic features of early B-type stars, including prominent hydrogen and helium absorption lines. Insets highlight these diagnostic transitions. Radial velocities were derived from ref.\cite{Xiang2022}, with $v_\mathrm{lsr}$ = -202 $\pm$ 5 km s$^{-1}$ for E-S1 (a) and $v_\mathrm{lsr}$ = -158 $\pm$ 1 km s$^{-1}$ for E-S2 (b). Rest wavelengths are indicated by gray dashed lines; magenta dashed lines mark the Doppler-shifted positions corresponding to each star’s systemic velocity.
}
    \label{fig:Extended_Data_Figure_3}
\end{figure}

\clearpage

\begin{figure}
\center
	\includegraphics[width=1\columnwidth]{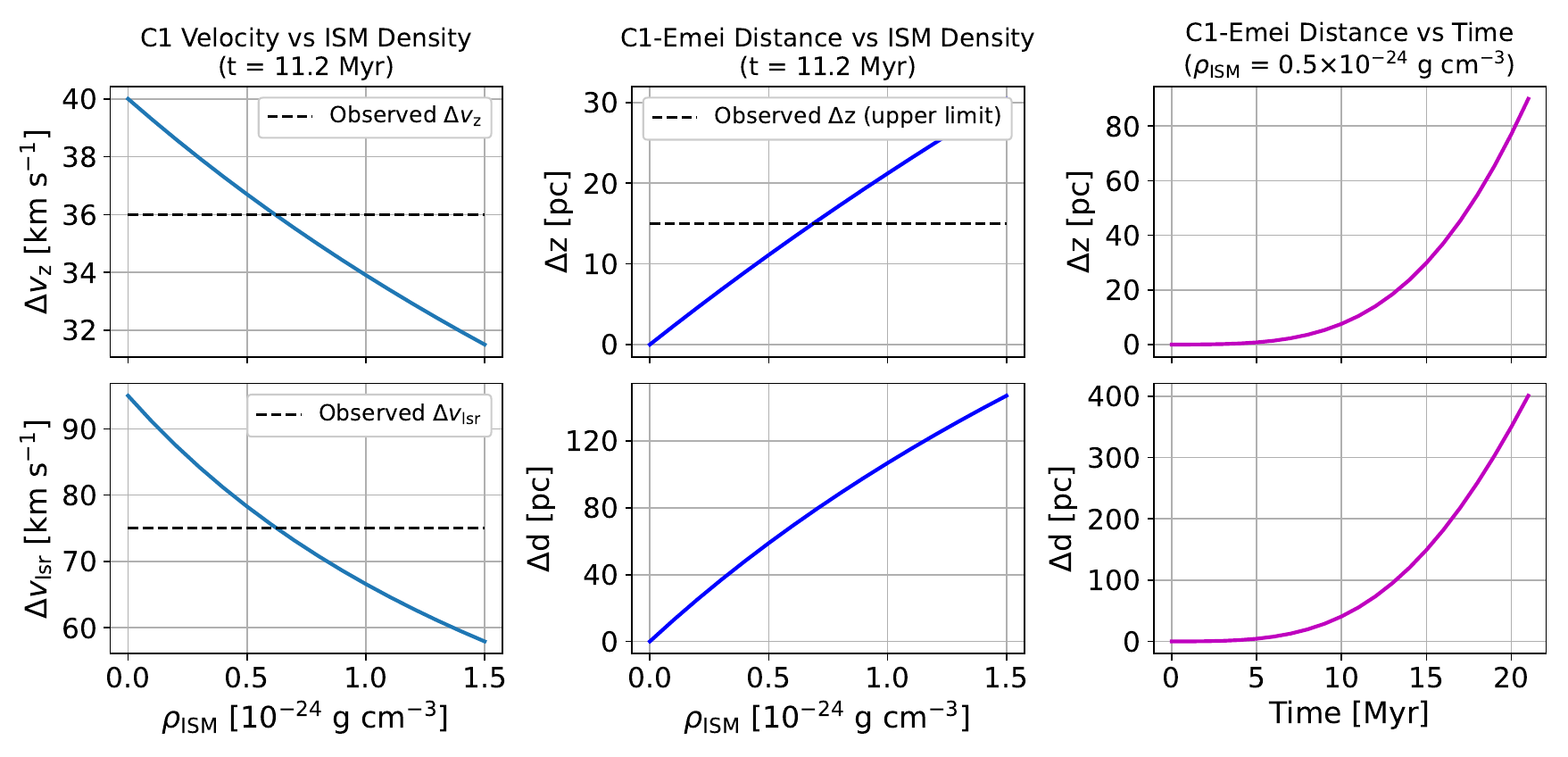}
   \caption{\textbf{Dynamics of C1 under different ISM conditions.} Top rows: results for initial vertical relative velocity 40 km s$^{-1}$; bottom rows: results for initial radial relative velocity 95 km s$^{-1}$. Left columns: velocity decrement as a function of ISM density at $t = 11.2$ Myr. Middle columns: separation distance from the Emei clusters as a function of ISM density. Right columns: temporal evolution of the separation distance at fixed ISM density ($0.5\times10^{-24}$ g cm$^{-3}$). Specifically, $\Delta v_{\mathrm{z}}$ and $\Delta v_{\mathrm{lsr}}$ denote the vertical and radial velocity differences between cloud C1 and the Galactic disk gas. $\Delta z$ and $\Delta d$ are the vertical and radial spatial separations between C1 and the star clusters. The dashed lines represent the observed relationships between these velocity and separation (upper limit) differences.}
    \label{fig:Extended_Data_Figure_4}
\end{figure}

\clearpage

\begin{figure}
\center
	\includegraphics[width=1\columnwidth]{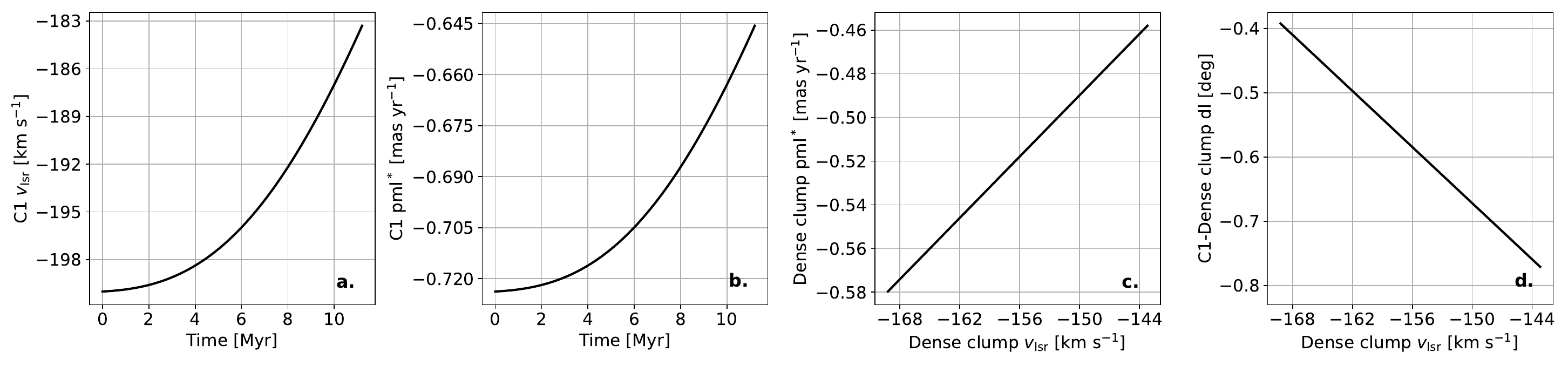}
   \caption{\textbf{C1 and dense clump kinematics.}
(a) Time evolution of C1 $v_{\mathrm{lsr}}$ under $\rho = 0.5 \times 10^{-24}$ g cm$^{-3}$ drag. 
(b) Corresponding pml$^*$ evolution. 
(c) Dense clump pml$^*$ vs. $v_{\mathrm{lsr}}$. 
(d) Predicted C1-clump angular separation (dl) vs. clump $v_{\mathrm{lsr}}$.}
    \label{fig:Extended_Data_Figure_5}
\end{figure}

\clearpage

\begin{figure}
\center
	\includegraphics[width=1\columnwidth]{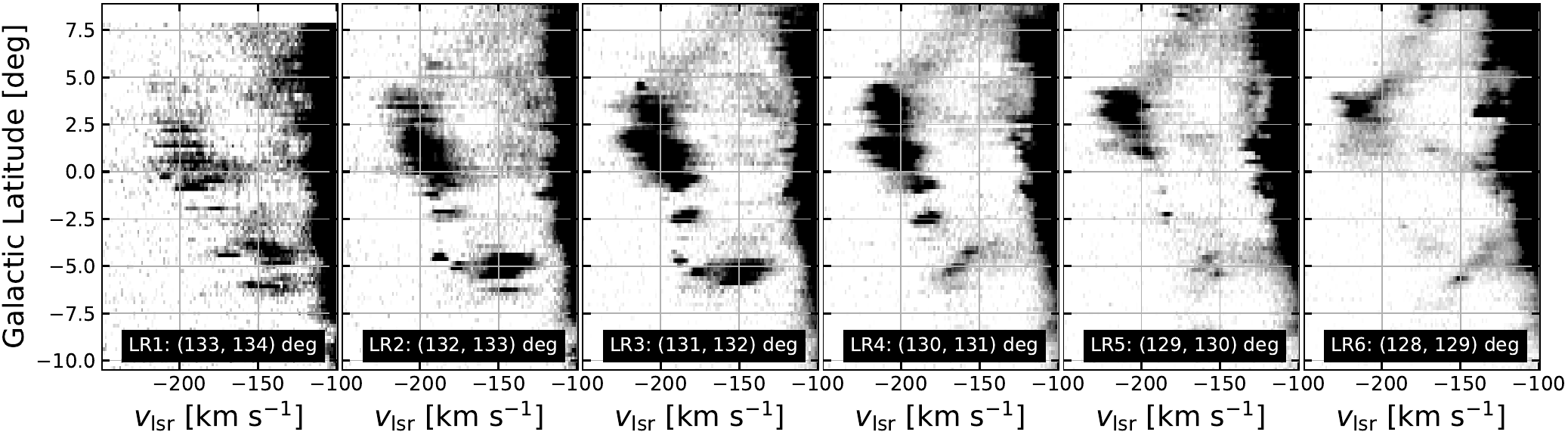}
   \caption{\textbf{The SFS signature of Complex H.} This latitude-velocity diagram uses EBHIS data from longitudes 134$^\circ$ to 128$^\circ$. 
The longitude range (LR) of each panel is indicated below it. 
For clarity in displaying diffuse structures, each subpanel is displayed at an individual depth range from 50\% to 90\% of its respective intensity levels.
The diagram reveals the ``$<$'' pattern of velocity gradient. 
This pattern shows a smooth connection between the Complex H components and the disk gas at $v_\mathrm{lsr}$ $\sim$ -105 km s$^{-1}$. 
This connection indicates the cloud's outer layers are merging with the Galactic disk and reveals active interaction.}
    \label{fig:Extended_Data_Figure_6}
\end{figure}

\clearpage

\begin{figure}
\center
	\includegraphics[width=.9\columnwidth]{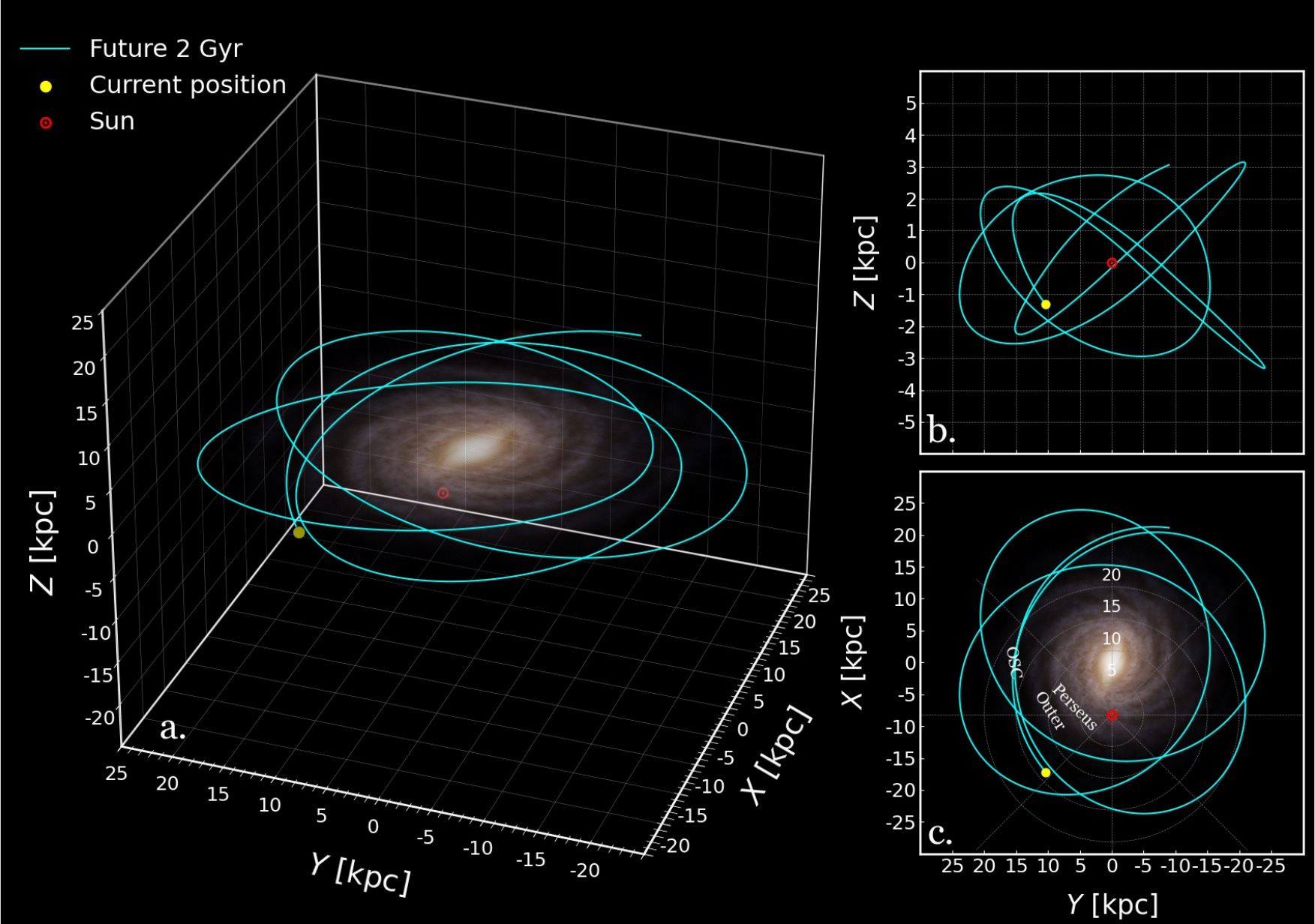}
\caption{\textbf{Galactic orbit reconstruction of the Emei clusters.} Orbital integration was performed using $Galpy$ with the MWPotential2014 gravitational potential \citep{Bovy2013,Bovy2015}, based on the clusters’ proper motion and the parent cloud’s systemic velocity. (a) Three-dimensional Galactic orbit overlaid on a schematic Milky Way disk illustration (Image Credit: ESA/Gaia/DPAC, Stefan Payne-Wardenaar).
(b) Edge-on and (c) face-on projections of the orbits, with spiral arm names labeled in the latter. The orbital path suggests SCH may be interacting with either the extension of the Outer Scutum-Centaurus (OSC) arm or the region beyond this spiral arm in the second Galactic quadrant, with the Emei clusters expected to cross the Galactic disk near this region.}
    \label{fig:Extended_Data_Figure_7}
\end{figure}

\end{document}